\documentclass[ 12pt,
 amsmath,amssymb,
 aps,
]{revtex4-2}
\usepackage{graphicx}
\usepackage{float}

\usepackage{enumitem}
\usepackage{dcolumn}
\usepackage{bm}
\begin{document}
\title{Employing Typicality in Optimal Control Theory}
\affiliation{Sciences Department, Holon Academic Institute of Technology, 52 Golomb Street, Holon 58102, Israel}
\author{Aviv Aroch}
\affiliation{The Institute of Chemistry, The Hebrew University of Jerusalem, Jerusalem 9190401, Israel}
\author{Shimshon Kallush}

\author{Ronnie Kosloff}
\affiliation{The Institute of Chemistry, The Hebrew University of Jerusalem, Jerusalem 9190401, Israel}

	\begin{abstract}

Controlling the dynamics of quantum systems is a crucial task in quantum science and technology. Obtaining the driving field that transforms the quantum systems to its objective is a typical control task.
This task is  hard,  scaling unfavorably  with the size of Hilbert space. To tackle this issue we employ typicality to assist in finding the control field for such systems. To demonstrate the method we choose  the control task of cooling the fine structure states of the AlF molecule, from relatively high temperatures which results in large Hilbert space. Using quantum typicality, we demonstrate that we can simulate an ensemble of states, enabling a control task addressing simultaneously many states. We employ this method to find a control field for cooling molecules with large number of internal sates, corresponding to high initial temperatures.

\end{abstract}
\maketitle

	\section{Introduction}

Controlling quantum phenomena has been a goal in quantum physics and chemistry.	
Quantum control theory addresses this topic and has evolved rapidly over the last three decades \cite{glaser2015training}.  One of the main goals of quantum control theory is to establish a series of systematic methods to manipulate and control quantum systems.  Quantum control theory has been implemented in physical chemistry, atomic and molecular physics, and quantum optics. In much of quantum control theory, the controllability of quantum systems is a fundamental issue \cite{huang1983controllability}. Controllability concerns the existence of a control solution for a specific task. This problem has practical importance since it closely connects with the universality of quantum computation and the possibility of achieving atomic or molecular scale transformations.  For finite-dimensional quantum systems, the controllability criteria may be expressed in terms of the structure and rank of corresponding Lie groups and Lie algebras \cite{albertini2003notions,ramakrishna1995controllability}. This method allows for a mathematical treatment of the problem in the case of closed quantum systems. 

The existence of a controllable task does not hint at how to obtain the control field.  
Iterative schemes  have been developed based on constrained
optimization has been developed \cite{peirce1988optimal,kosloff1989wavepacket,palao2003optimal,serban05,gross07,eitan2011optimal,reich2012monotonically}.
Controlling systems with Large Hilbert space is a challenging problem mainly due to its computational complexity. 

The computational effort measures the complexity of the problem we are trying to solve. In our case, the problem is twofold; our primary step requires solving the time-dependent Schr\"odinger equation for each state. The complexity scales between $M \times N \log N$ and $M \times N^2$ where $N$ is the size of Hilbert space of our system and $M$ the number of time-steps which in turn scale as $M \propto O(\Delta E \times t)$ where $t$ is the time interval and $\Delta E$ the energy range \cite{kosloff1988time}. The number of iterations for solving a unitary control problem scales like $K!$ where $K$ is the size of the transformation \cite{palao2003optimal}. The complexity class of the cooling transformation, which addresses $K$ states simultaneously,
is not known and could be between polynomial in $K$ and factorial $K!$. 
Altogether the control problem is computationally highly complex \cite{arora2009computational}.

This issue requires establishing methods that reduce the number of states $K$ needed for optimization and the complexity. To achieve this task, we will employ the properties of quantum typicality. 

Quantum typicality states that a single quantum state can typically well describe local expectation values of a quantum system. 
This statement applies to Schr\"odinger type dynamics in high dimensional Hilbert spaces. As a consequence, individual dynamics of expectation values converge to the ensemble's average \cite{bartsch2009dynamical}. We will employ quantum typicality by using random phase wavefunctions for control problems. 

The random phase wave function method uses an ensemble of pure states, which creates an efficient representation of the mixed state of the entire system. The convergence of the RPWF method becomes faster as the size of the Hilbert space (and the number of random phases) increases, namely with the rise of initial temperature.

The present study aims to develop an optimal control algorithm to cool the molecular internal degrees of freedom, a multi-state control task.
The control objective is to increase the system's purity ${\cal P}=Tr\{\hat \rho^2\}$. Alternatively, cooling can be defined as lowering the effective temperature $T_{eff}$
defined by the von Neumann entropy of the ensemble \cite{von2018mathematical}. $T_{eff}$ is the temperature of a passive Gibbs state with the same von Neumann entropy \cite{uzdin2018global}.

The field of cold and ultra-cold molecules has rapidly grown in the last decade. Cold molecules have an essential role in many active areas in science amongst tests of fundamental physics, cold chemistry, quantum technologies (QT), and quantum information (QI)\cite{krems2009cold,carini2015enhancement,kallush2008unitary,yeo2015rotational,lien2014broadband}. 

There are many ways of cooling particles; one of the most effective is radiative cooling.
The basic technique is to locate a closed loop of stimulated excitation and spontaneous emission. The cooling is achieved by entropy
removal by spontaneous emission. This technique has been employed to reduce the temperature of translational as well as internal degrees of freedom of atomic and molecular species \cite{aroch2018optimizing,itano1995cooling,zeppenfeld2012sisyphus}. 

A crucial consideration in cooling internal degrees of freedom in molecules is enforcing a close transition cycle.
Excitation may result in the molecule ending in a different state outside the closed-loop after decay; thus, cooling molecules by laser excitation is complicated, and only a few examples exist. However, the molecular complexity makes cold molecules applicable for a broader range than atoms. 

We choose optimal control theory to overcome the molecular complexity and enforce closed-loop solutions.
Controlled laser fields are employed to remove frequencies that damage the required transition. Control can steer a quantum system from its initial state to its final one (target state). Optimal Control Theory (OCT) was created to do this task with maximum fidelity (defined by the user). 

To control a specific system, we first need to describe the state and its evolution in time. In our system, we can define the rotational state employing the eigenstates of a rigid rotor as a complete basis. The state is propagated as an isolated quantum system evolving by a unitary transformation.

Optimal Control Theory (OCT) is employed for an isolated system to obtain the field leading to the target state. However, this alone will not lead us into a colder state (pure state); entropy is invariant under unitary transformation. Therefore cooling requires dissipation that can change the entropy of our quantum system. 

Assuming that the target of the process is to reach a pure, single state, any proposed mechanism for the process has to maintain the population of the single-target state while allowing the population of all other states to repopulate selectively. However, as was shown by \cite{maday2003new,reich2013cooling}, the speciﬁc choice of the precooling transformation is a subtle issue. As implied by the ergodic theorem \cite{brixner2004quantum}, for any initial state under multiple cycles of a given ﬁeld-driven unitary transformation and subsequent decay, the ﬁnal state will be the invariant under the whole transformation. After many excitation-relaxation cycles, the memory of the initial state will be erased. The entire transformation can be described by the following, 
let ${\cal U}$ be   the unitary super operator and  ${\cal D}$ be dissipative super operator; then there is a stationary state $\hat \rho_{SS}$ that will obey 
 \begin{equation}
    \lim _{n \rightarrow \infty}\left({\cal U} {\cal D} \right)^{n} \hat \rho = \hat \rho_{ss}
\label{Full-map}
\end{equation}
where $\hat \rho$ can be any initial state, and ${\cal U}\bullet = \hat U \bullet \hat U^{\dagger}$. That is, under a given ${\cal U}$ and ${\cal D}$, the system will ﬁnally evolve from any state into the single stationary $\hat \rho_{ss}$. In this way, we look for a unitary control transformation that under a known dissipator will lead into a purer state \cite{aroch2018optimizing}. 

The ultimate goal of molecular cooling is to reach a
large sample of pure state molecules that can be transformed to a BEC. 
This task can only be achieved in stages.
The initial sample should be as pure as possible, pure from other isotopes, other quantum states (hotter ones in our case), etc.,  is crucial. 
A crucial intermediate step is to obtain a molecular 
Magnetic Optical Trap (MOT). 
For achieving this step, an ensemble of a single rovibrational state should be created.

Our control objective is designed to enhance the initial preparation; it can be used as a purification method to add molecules that are not resonant with MOT transition to be added
and thus increase the number density of the initial ensemble.

Specifically, we choose to cool the AlF molecule rotationally. Spectroscopic measurements and detailed analysis \cite{truppe2019spectroscopy}, have shown that such a task is feasible using optimal control theory. This molecule belongs to a family of which the vibrational manifolds are closed due to a Frank-Condon
coefficient close to one.
Our obstacle is the large initial number of rotational states.

The present theoretical study aims to obtain a high fidelity control task with low computational effort.  We have implemented this method on the fine-structure levels of the AlF molecule with a large total orbital angular momentum at 30K. We will employ optimal control theory to ﬁnd broadband-shaped pulses, steering the system into a colder state. 

The paper is arranged as follows: Section II  describes the model in which we implemented our tools, in section III describes the theoretical tools used for achieving control. Section IV presents the results, which are discussed and summarized in the concluding Sec. V.

\section{The Model}
\label{sec:model}

The model we employ is the rovibrational structure of the AlF molecule. The state  of the systems is defined by the combined density operator of two electronic surfaces:

\begin{equation}
\\{\hat \rho}  = {\hat \rho _g} \otimes {\hat P_g} + {\hat \rho _e} \otimes {\hat P_e} + {\hat \rho _c} \otimes {\hat S_ + } + \hat \rho _c^* \otimes {\hat S_ - }\
\end{equation}

Where $\hat P_{g / e}$ are the projection operators of the ground and excited electronic state, $\hat{S}_{+/-}$ are the electronic raising and lowering operators, $\rho_{g / e}$ are the density operator for the rovibrational ensemble within the ground and excited electronic states, and $\hat \rho_{c}$ is the density operator of the nuclear coherence between the surfaces. 

For AlF, we assume an initial temperature of $\sim 30^0 K$, for which the the population of the states occupies up to $J = 11$ (144 sub-levels).

The Liouville von Neumann equation governs the evolution of the system:
\begin{equation}
\frac{{d\hat \rho }}{{dt}} =  - \frac{i}{\hbar }[\hat H,\hat \rho ] + {\cal L}_D(\hat \rho )
\label{eq:liouvill}
\end{equation}
The first term is the coherent dynamical part governed by the Hamiltonian and the second is the dissipative part of the dynamics describing spontaneous emission. This equation represents the dynamics of an open quantum system. 

For optical transitions with multiple pulses, there is a distinct timescale separation between the light-induced step, which occurs in less than a  picosecond and is unitary, the incoherent decay which occurs in tens of nanoseconds, and the pulse repetition rate, which is in the $MHz$ to $KHz$. 
Each cooling cycle could be separated within this picture into two parts: (1) The short-time interaction of the external field and the molecular system. Since this step is unitary, the density operator can be decomposed to energy eigenstates, and each component can be computed in a wave function framework. Then (2) a slow and field-free, spontaneous decay takes place. In this step, the coherences developed between energy eigenstates during the laser-controlled stage are erased.

The Hamiltonian, which governs the unitary part of the dynamics, can be written as :
 \begin{equation}
\begin{array}{l}
{\hat H_t = {{\hat H}_0}} + {{\hat V}_t}\\
{{\hat H}_0} = {{\hat H}_g} \otimes {{\hat P}_g} + {{\hat H}_e} \otimes {{\hat P}_e}.
\end{array}\
 \end{equation}
Where ${\hat H}_{g/e}$ is the ground and excited rotatioanl Hamiltonian.
The interaction of the system with light, assumed to be linearly polarized to the lab $z$ axis is described by ${\hat V}_t$:
\begin{equation}
{{\hat V}_t} =  -  \hat \mu_z  \otimes ({{\hat S}_ + }\varepsilon_z(t)  + {{\hat S}_ - }{\varepsilon_z(t) ^*})
\end{equation}

Where $\hat \mu_z$ is the transition dipole moment along the $z$ spatial direction and $\varepsilon_z(t)$ represents the time-dependent field along the same direction.  

For mildly cold temperatures  ($T_{initial} \le 30K$) and high vibrational frequency, we can assume that the molecules are initially in their ground $v=0$ state. ${\hat H_{g/e}}$ are the field-free rotational Hamiltonians for the ground vibrational state. Moreover, for the $AlF$ and chemically similar molecules, vibrational excitations in the electronic transition are negligible due to the highly restricting FC factors \cite{truppe2019spectroscopic}. As a result the modeling is thus restricted to $v=0$. 

Under the Hund's case $a$, applicable to our case, the rotational states of the model are expanded by the symmetric top basis \cite{zare1988angular}
\begin{equation}
    |J ,\Omega M\rangle=\left[\frac{2 J+1}{4 \pi}\right]^{\frac{1}{2}} D_{M,\Omega}^{J}(\phi, \theta, 0)
\end{equation}
Where $J$ is the total molecular angular momentum, $M$ and $\Omega$ are projections on the spatial (Z) and molecular (z) axes, respectively. Here, $D_{M \Omega}^{J}$ is the rotational tensor.
For the ground electronic ${ }^{1} \Sigma$ state, the projection of the spatial electronic angular momentum on the molecular axis is 0 (L=S=0), therefor $\Lambda=0$, and J=N=R. The rotational Hamiltonian becomes 
\begin{equation}
    \hat{H}_{\text {rot}}(r)=B \hat{\boldsymbol{R}}^{2} = B \hat{\boldsymbol{J}}^{2} 
\end{equation}
where $B$ is a rotational constant, 
and $\hat R$ is the nuclear rotational angular momentum operator, which is equal to
${\boldsymbol{R}}={J}-{L}-{S}$
, L is the electronic orbital angular momentum and S is the electronic spin angular  momentum.

$^1\Sigma$ energies are then
\begin{equation}
    E(^1{\Sigma} ; J) = {B}J(J+1)
\end{equation}
and its corresponding eigenstates can be defined by $|J\Omega M\rangle$ when $\Omega=0$.
\newline
In the excited $^1\Pi$ state $\Lambda=\pm 1$ S=0, and thus $\Omega= \pm 1$.

The rotational Hamiltonian for the excited state, $^1\Pi$, is 
\begin{equation}
    \hat{H}_{\text {rot}}(r)=B \hat{\boldsymbol{R}}^{2}=B(\hat{\boldsymbol{J}}-\hat{\boldsymbol{L}})^2
\end{equation}
its energies are 
\begin{equation}
    E(^1{\Sigma} ; J) = {B}[J(J+1)-1]
\end{equation}
the corresponding eigenstates are $|J ,\Omega M\rangle$ where $\Omega= \pm {1}$.

The transitions between the two electronic states ${ }^{1} \Sigma \rightarrow{ }^{1} \Pi$ are dictated by dipole selection rules $(\Delta J=0,\pm 1)$, denoted as R,Q,and P branches,respectively.The coupling elements can be found by calculating the overlap of any two eigenvectors with the dipole operator,
\begin{equation}
    \begin{aligned}
\int D_{m^{\prime} \Omega^{\prime}}^{j^{\prime}}(\theta, \phi, 0) D_{0 q}^{1}(\theta, \phi, 0) D_{m \Omega}^{j}(\theta, \phi, 0) d \Omega \\
&=8 \pi\left(\begin{array}{ccc}
j & 1 & j^{\prime} \\
-m & 0 & m^{\prime}
\end{array}\right)\left(\begin{array}{ccc}
j & 1 & j^{\prime} \\
-\Omega & q & \Omega^{\prime}
\end{array}\right)
\end{aligned}
\end{equation}
where $\mu_{0 q}$ is the transition dipole moment proportional to $D_{0 q}^{1}(\theta, \phi, 0)$. The value of q is determined for a given transition case and is
equal to $q=\Omega-\Omega^{\prime}$.

At thermal equilibrium, the initial state $ \rho_{e q}$ is characterized by thermally distributed  in the quantum rotational states
 \begin{equation}
     \hat{\rho}_{e q}=\frac{1}{Z} \sum_{j} e^{-\beta E_{g}^{j}}|J 0 M \rangle\langle J 0 M|
 \end{equation}
where $\beta=1 / k_{b} T$ and the sets 
$|J \Omega M \rangle$  and  $\left\{E_{j}\right\}$
are the eigenstates and eigenenergies of the system, and Z is the partition function.

The dissipating part of the dynamics is generated by the Liovillian super operator ${\cal L}_D$, Eq. (\ref{eq:liouvill}).
Integrating in time leads to the transition map $\Lambda_t=e^{ {\cal L}_D t}$.
Assuming that the timescale between pulses is longer than the spontaneous emission, the transition map ${\cal D} $ of the spontaneous emission can be defined. The matrix elements of ${\cal D}$ are from a given excited state energy eigenstate to the ground state manifold of states.
They can be calculated employing Fermi's golden rule:
\begin{equation}
\Gamma_{i \rightarrow f}=\frac{2 \pi}{\hbar}\left|\left\langle f\left|\mu\right| i\right\rangle\right|^{2} \rho\left(E_{f}\right)
\label{decay}
\end{equation}

where $\left\langle f\left|H^{\prime}\right| i\right\rangle$ is the matrix element of the electronic transition dipole  between the final and initial states, $\rho\left(E_{f}\right)$ is the density of states in vacuum (number of continuum states divided by $d E$ in the infinitesimally small energy interval $E$ to $E+d E$ ) at the energy $E_{f}$ of the final states. With  Eq. (\ref{decay}) we have defined the decay rate ($\sim 10^{-6}~ second$), it is important to note that because the rotattional selection rules ,in our system, dictates narrow band transitions (all transitions are almost equal), $\omega^3$ can be neglected. Assuming that we wait sufficient time and normalize it; we get the decay probability of our system.

It is important to note that the coherent step yields only unitary transformations and therefore does not change the purity when the latter decay step is known for the loss of purity and thus changes the system's temperature. 

We can use the defined Full map transformation Eq. (\ref{Full-map}) and get a final state that is invariant under the whole transformation. 
After many excitation-relaxation cycles, the memory of the initial state will be erased and the system will finally evolve  into the single stationary $\rho_{ss}$.

One can obtain the state $\rho_{ss}$ by diagonalization of Eq. (\ref{Full-map}). The eigenstate then gives the stationary state with a unit eigenvalue, while the next eigenvalue indicates the system's convergence rate to the steady-state. 

Using the ergodic theory compels the system to be closed. Since  ${\cal D}$ is fixed, the task is to find the unitary transformation ${\cal U}$ that is will lead to the designed stationary state. 

To associate an effective temperature to the state, we employ the von-Neuman entropy to scale the purity and define the effective temperature. The idea comes from information theory, where the entropy is related to the probability distribution of an ensemble \cite{von2018mathematical}.  The entropy is defined as:
\begin{equation}
  {S_{VN}} =  - tr\{ \hat \rho \ln \hat \rho \}  \le  - \sum\limits_j {{P_j}} ln{P_j}
  \label{SVN}
\end{equation}
\newline
where $\hat \rho$ is the system's density matrix, and $P_j$ is the probability to be in the energy eigenstate $j$. This is the only contribution to the entropy, assuming that quantum coherences do not survive the spontaneous emission incoherent step. Equality will be obtained when the system is diagonalized in the energy domain. It is important to note that entropy is invariant under unitary transformation. Therefore any steady-state reached after cooling can be transformed by unitary transformation to a passive state with the same entropy \cite{lenard1978thermodynamical}.
To define a temperature of any non-thermal state, we associate it
with the temperature of a thermal state with the same vN entropy \cite{uzdin2018global}.

\section{Methods}

\subsection{Optimal Control Theory (OCT)}
 Quantum Optimal Control Theory (OCT) is a branch of coherent control, a quantum mechanical based method for controlling dynamical processes. The basic principle is to control quantum interference phenomena typically by shaping the phase of laser pulses \cite{tannor2007introduction, werschnik2007quantum,peirce1988optimal,kosloff1989wavepacket}.
OCT is formulated as a maximization problem, and seeks a time dependent field that maximizes the expectation value of an operator in final time. 
 
Consider a quantum system in an initial state: ${\hat \rho_0 } = \sum\limits_{k=1}^N p_k | \psi_k^0 \rangle \langle \psi_k^0| $, where the set $\{ \psi_k^0 \}$ is energy eigenstates of the system, and $N$ the size of Hilbert space.
The control will seek a field that maximizes the expectation value of the operator $\hat O$ at final time T:
\begin{equation}
  {J_{\max }}(\varepsilon ) \equiv \sum\limits_{k = 1}^N {p_k\langle {\psi _k}(T)|\hat O\left| {{\psi _k}(T)} \right\rangle } 
  \label{eq:j}
\end{equation}
where ${\Psi _i}(T)$ describes the state that results from the interaction of the system with the field $\varepsilon$at the final time $T$. 
The governing of the dynamics of the system by the Schr\"{o}dinger equation $i \frac{\partial}{\partial t} |\psi \rangle = {\hat H} |\psi \rangle$. The quantum dynamics is enforced by adding an additional cost term to the functional, according to the Lagrange-multiplier method:
 \begin{equation}
{J_{con}} = \sum\limits_{k = 1}^N { - 2{\rm{Re}}\int\limits_0^T {\langle {\chi _k}(t)|\frac{d}{{dt}} + i\hat H(t)} \left| {{\psi _k}(t)} \right\rangle dt} 
\label{eq:j2}
 \end{equation}
where $\{\left\langle{\chi_k} (t)\right|\}$ are the set of time dependent Lagrange function multipliers. 
To regularize the solution with a limitation over the intensity another penalty term to the functional is added \cite{degani2009quantum}:
\begin{equation}
   {J_{penal}}(\varepsilon ) =  -  {\alpha }{\int_0^T {\varepsilon^2(t)dt} } 
\end{equation}
where $ \alpha $ is a scalar Lagrange multiplier. 
 
The overall object of maximization is the following functional:
\begin{equation}
J = J_{max}+ J_{penal} + J_{con}
\end{equation}
The maximization of the fitness $J$ is the control task.
Functional derivatives with respect to the various field are then taken resulting in the following system of equations:

Each of the set of the $|\chi_k(t) \rangle$ Lagrange multipliers will obey a time reversed Schr\"{o}dinger equation:
 \begin{equation}
    \frac{d {\left\langle { \chi _k}(t)\right|}}{dt} = i\left\langle {{\chi _k}(t)} \right|\hat H(t) 
\label{revsch}
 \end{equation}
with the boundary conditions: $|\chi_k(T) \rangle= \hat O | \Psi_k(T)\rangle$ \cite{palao2002quantum}.
 
The Krotov iterative method is applied to obtain a monotonic growth of of the fitness $J$ at each iteration with the updated field so that:
  \begin{equation}
    \varepsilon^{l + 1}(t) = \varepsilon^{l(t)} - \frac{1}{\alpha }{{\sum\limits_{k = 1}^N {{\mathop{\rm Im}\nolimits} \left\langle {\chi _k^l(t)} \right|\hat {\mu} } \left| {\psi _k^{l + 1}(t)} \right\rangle } }
 \label{fieldcor}
 \end{equation}
where $\bullet^{^{l}}(t)$ denote the quantity after the the field after the $l-th$ iteration. 
 
Note that the scheme of Eqs. (\ref{revsch}) (\ref{fieldcor}) is somewhat similar to the simultaneous optimization scheme that is required for unitary transformations and quantum gates. However, for cooling each cycle at the final transformation erases the relative quantum phase between the various optimized set initial states. This leaves the resulting fitness measurement at the level of classical transition probability between the initial and final state and removes the need to evaluate quantum phases.
 
For large systems, the convergence of OCT is fast in the first iterations but later saturates or becomes stagnant. Therefore, accurate solutions for large systems of the control problem are difficult to reach. A good initial guess for the control field can speed convergence considerably. Ideally, a control field of an easy-to-solve small quantum system could serve as an initial pilot guess for large quantum systems \cite{kallush2008unitary}. 
 
The first step is to find the pilot field for which we seek a
solution to the control problem of dimention $K$ for a small number of random phase states $L$. We will then explore the universality of the control field for the increasing number of states. 
We will use optimal control theory, based on Krotov’s algorithm, which guarantees monotonic convergence \cite{tannor1992control}, to do so.
 
The difficulty in converging a control field to generate state-to-state transitions can be related to the algebraic structure of the control Hamiltonian. We have seen that when the initial or target states are superpositions of generalized coherent state, no relation exists between the control fields of such targets for different Hilbert space sizes.

\subsection{Quantum Typicality}

Typicality describes a property of a system where a typical state can present  an assembly of similar states.  This set of states should have a narrow distribution of some feature (e.g., drawn according to the same distribution, sharing the same energy, etc.) and therefore yield a very limited distribution of expectation values. The typical state will fit the expectation value of the complete set of states. 

Quantum typicality was first noted by Schr\"odinger and von Neumann when they were trying to incorporate statistical mechanics with quantum mechanics.  They inferred that the wavefunction of a complex system can
have statistical properties.

In their approach, when discussing thermalization in isolated quantum systems, one should focus on physical observables instead of wave functions or density matrices describing the entire system. This approach is similar to Eigenvalue Thermaliziation  Hypothesis (ETH), in which the focus is put on macroscopic observables and “typical” configuration. 

Eigenvalue Thermaliziation  Hypothesis (ETH) implies that the expectation values of local observables and their fluctuations in isolated quantum systems  relax to (nearly) time-independent values that can be described using traditional statistical mechanics ensembles. This has been verified in several quantum lattice systems and, according to ETH, should occur in generic many-body quantum systems. ETH states that the eigenstates of generic quantum Hamiltonians are “typical” in the sense that the statistical properties of physical observables are the same as those predicted by the microcanonical ensemble. \cite{rigol2008thermalization,von2010proof,d2016quantum}

Our quantum typicality refers to an idea that anticipates that almost all quantum systems will have similar dynamical properties
\cite{reimann2007typicality}.

\subsection{Random states}

Controlling the dynamics of an extensive system is practically impossible when the system becomes large and complex. 
Formally our control strategy requires $N$ states Eq. (\ref{eq:j}) and (\ref{eq:j2}). Can we reduce the number of states to $ L < K < N$ ?
We know from Ref. \cite{romero2020equilibration} that sampling quantum states at random can be seen to be induced by the sampling of unitaries at random. Sampling a set of unitaries, which are bounded, means we sample a set of actions we can apply to our state, which is global.

We will now sample using  a random state,  as follows:  
\begin{equation}
    |\Psi({\vec{\theta}})\rangle=\frac{1}{\sqrt{N}} \sum_{j}^{N} e^{i \theta_{j}}\left|\phi_{j}\right\rangle
\end{equation}
where $\vec \theta =(\theta_1,\theta_2 ...\theta_N)$ is a vector
of random phases and $N$ is the size of the Hilbert space \cite{nest2007quantum,kallush2015orientation}.
Employing a random set defined by different $\vec{\theta}$ we can resolve the identity:
\begin{equation}
    \hat I = \lim_{ M \rightarrow \infty}\frac{1}{M}\sum_l | \psi(\vec \theta_l) \rangle \langle \psi(\vec \theta_l) |
\end{equation}
Employing now $K$ random states to sample the control and relying on  quantum typicality we expect $K < N$.

\section{Results}

Our primary goal in this study is to develop a method to control systems with large Hilbert space while minimizing the computational effort. The objective of the present model was to cool by increasing the system's purity at the final steady-state after multiple cycles using the OCT algorithm. It is important to note that this objective is a multi-state problem.
We have shown that such a task is possible \cite{aroch2018optimizing}, but the main drawback is the computation scaling of the problem with the system's size.

Assuming that the target of the process is to reach a colder state, any proposed mechanism for the process has to maintain the population of the target state while allowing the population of all other states to repopulate selectively. 
After many excitation-relaxation cycles, the memory of the initial state will be erased, and the obtained transformation will be of which mentioned in Eq. (\ref{Full-map}). In this model, rotational cooling of AlF Sec. \ref{sec:model}
we define the cost function $J$ to populate states which will
through spontaneous emission will populate lower $j$ values
and penalize increase in $j$ values.
To control this type of system we have created several realizations for an increasing number of random states $L$ to  control the transformation.
We wanted to check:

\begin{enumerate}
\item
Does the total transformation can be represented.
\item
If it does, how many states $L$ are required to converge the entire system and the cooling transformation sufficiently?  (where $K$ is the size of the cooling transformation).
\end{enumerate}

The solution we obtained from the OC algorithm (the control field) was employed as a pilot field for the following realization. We hoped to decrease the complexity of the problem by reducing the amount of iteration needed to solve the control problem for the next realization. 

Employing the control field from a small $L$, as a pilot field, to a larger one did not significantly improve the convergence. Such a behavior has been observed for generic quantum objectives \cite{kallush2011scaling}.

\begin{figure}[H]

\includegraphics[scale=0.5, angle=0]{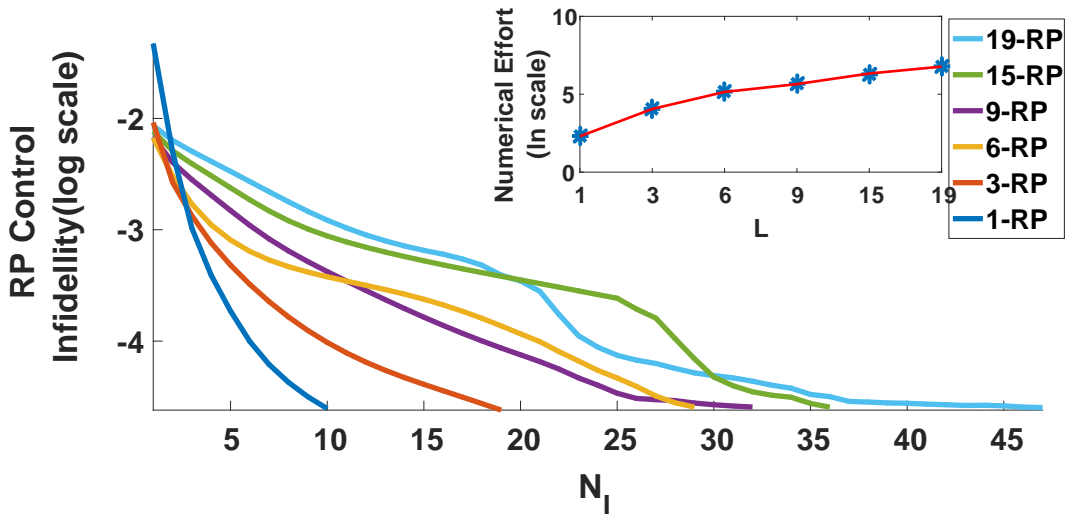}
\caption{\label{fig:RPControlVsNI}(main) The controlled Infidelity (log scale), of different number random phase sampling, as a function of the number of iterations required to converge the control($\approx{99\%}$ at the target state). The number of RP wavefunctions is assigned by different colors in the legend. 
The number of states of each model influences the number of iteration, increase monotonically. {Inset} The numerical effort (log scale) as a function of the number of RP states. The numerical effort is a function of Hilbert space times the number of iterations, showing a polynomial scaling.  }
\end{figure}

Figure \ref{fig:RPControlVsNI} displays the control fidelity of systems with a different number of random states $L$ as a function to the number of iterations. To estimate the number of states needed to converge the control, we have sampled a small set and checked the obtained control field on the full-sized (Hilbert space) dynamics. 

We have a controlled scheme of a single field polarized to the $Z$ spatial axis. The initial population distributed among different random states corresponds to a system with $J = 11 \to N = 430$. The target was to move the population to all states in Hilbert space, up to excited state at $J\leqslant10$ and eliminate any transitions to higher $J$ states that open decay channels to even higher states.  The control field can manipulate the system towards the target state with a fitness of $99\% $. However, this fidelity fits only the contracted space; later on, we will show that the same field acting on the full space system results in lower fidelity. Thus, some impurities are leaking in.

In the inset \ref{fig:RPControlVsNI} we show how the Numerical effort increases polynomially. The effort is defined by the following: ${N_{Effort}}=\ln ({N_I} \times {M_{RP}})$ where $M_{RP}$ is the number of random phase states used in this model and $N_{I}$ states the number of iterations needed for the OC algorithm to converge to the predetermined thresh hold.

\begin{figure}[H]
\includegraphics[scale=0.5, angle=0]{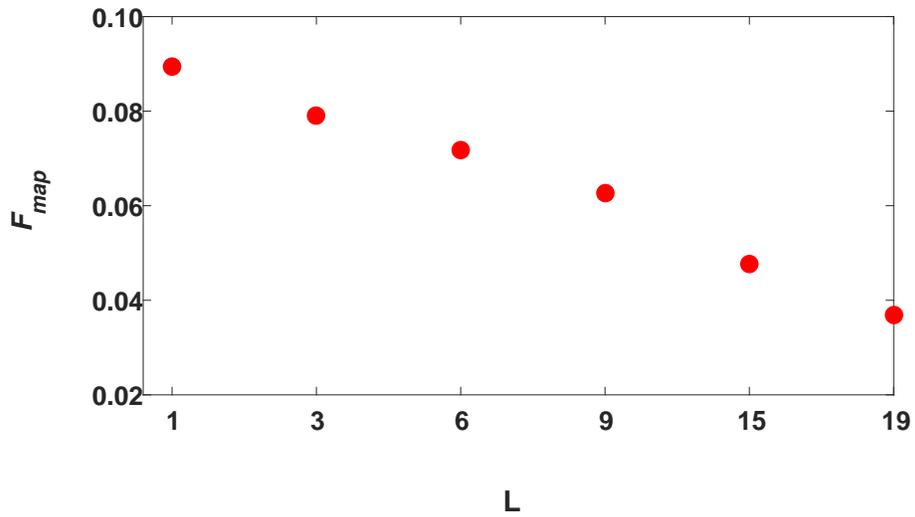}
\caption{\label{Full state Fidelity}
The infidelity as a function of the number of random phase base functions $L$.
The infidelity is calculated for the full transformation.}
\end{figure}

As mentioned, we have used the controlled field on the full system; figure \ref{Full state Fidelity} shows the infidelity of the entire system as a function of the number of random phase states $L$. To  check the quality of the transition we used the field from the controlled scheme on the dynamics of the entire system. We find a monotonic growth of fidelity corresponding to the number $L$ of random states. 

The fidelity of the full transformation as expected is smaller than  the one obtained from the control calculated from a finite set of random phase states. To find the number of random states which lead to convergence, we extrapolate our results, taking its infidelity as a measure.

\begin{figure}[H]

\includegraphics[scale=0.5, angle=0]{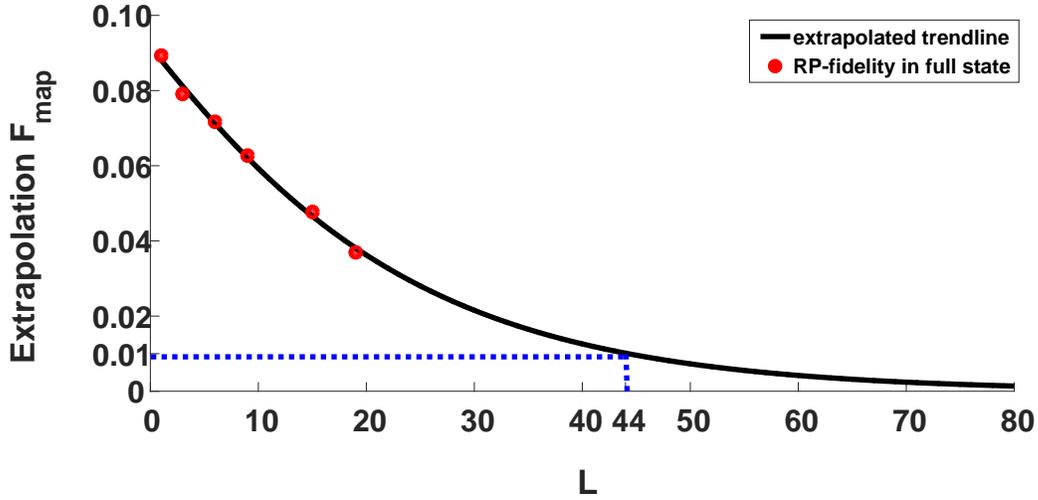}
\caption{\label{fig:extrapolate} Fitting the infidelity of the full transformation (red dots) as a function of the number of RP states (blck curve). Extrapolating to high fidelity allows to estimate the effective size of the cooling transformation (44,0.01).}
\end{figure}

In figure \ref{fig:extrapolate} we have fitted our calculated points to an exponential curve.
Defining an acceptable threshold we get the number of sampling states $K$ required to converge  the full  system of dimension $N$. 

The marked point is  a good guess to the complexity of the transformation corresponding to an effective transformation
size of $K=44$. Nevertheless, the computation effort for this sampling space is vary high. The effective transformation size is still much smaller than the dimension of the full transformation.   

A concrete measure for cooling in our context, is the change in normalized entropy.
Employing Eq. (\ref{SVN}) we define the normalized entropy decrease:
\begin{equation}
    \Delta {S_{eff}} = \frac{{S_{FS}^{RP} - S_{initial}^{J = 11}}}{{S_{Th }^{J = 10} - S_{initial}^{J = 11}}}
    \label{Effective_S}
\end{equation}
where $S_{FS}^{RP}$ is the entropy of the full system obtained from the control sequence, $S_{initial}^{J = 11}$ is the thermal entropy of the  initial state (J=11), and $S_{Th }^{J = 10}$ is the thermal entropy of  target for our transformation.  The normalization is with respect to the thermal entropy difference between $J=11$ to $J=10$ (at T= $\sim 30^0 K$).

\begin{figure}[H]
\centering

\includegraphics[scale=0.5, angle=0]{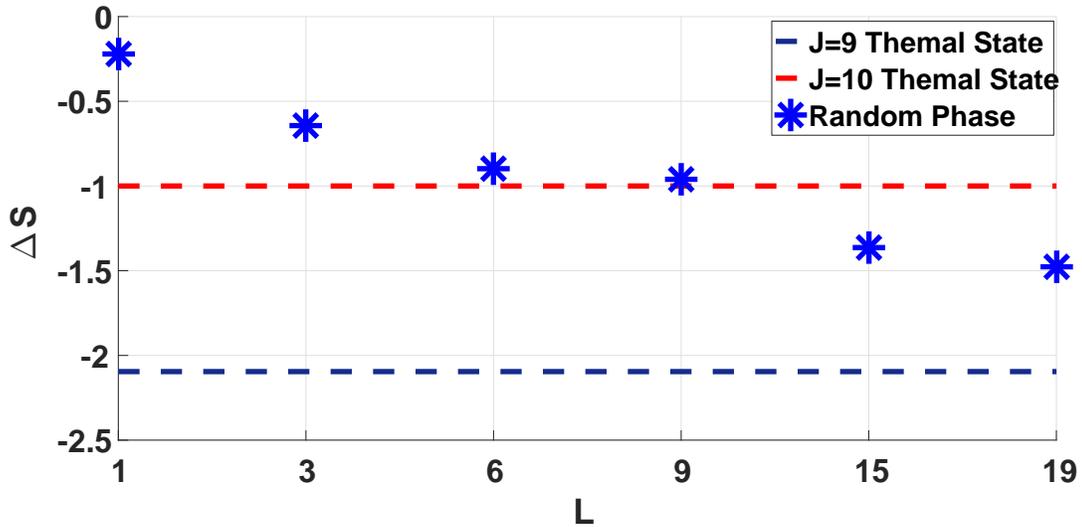}
\vspace{-0.5in}\caption{\label{fig:deltaS} The monotonic decrease of normalized entropy, Eq. (\ref{Effective_S}) with $L$, for the complete transformation obtained by each RP model (blue stars).
The entropy value of the thermal state at J=10 is marked by a red line and  for J=9 by a blue line. The objective of full  transformation was to cool from $J=11$ to J=10, but it  is clear  that for the 19-RP model we accomplished an even colder state.}
\end{figure}

Figure \ref{fig:deltaS} shows the effective change in entropy for each model (RP). A monotonic decrease of the effective entropy is a clear indication for cooling. 

The samples describe different realisation, where the initial state were created randomly. We have used these realisations for the same transformation and used it on the full spaced system. The fidelity we have got on the full system with respect to the desired transformation as calculated, and built for the calculation of the standard deviation, which is defined by $STD =\frac{\sqrt{<J^2>-<J>^2}}{|<J>|}$.
In the next figure we show how the STD behaves in different models.
\begin{figure}[H]

\includegraphics[scale=0.5, angle=0]{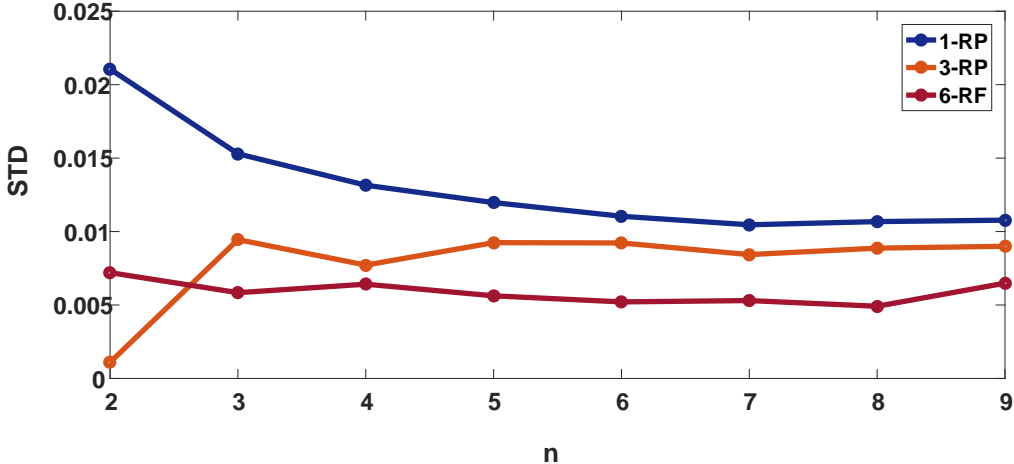}
\caption{\label{fig:STD} The Standard Deviation (STD) of the 
infidelity of the total transformation as a function of the number of samples ($n$).  The blue curve is data obtained by optimizing a single RP model. The orange curve was obtained for simultaneous transformation for 3-RP sampling. The red curve was obtained for simultaneous transformation for 6-RP sampling.}

\end{figure}

An ensemble of typical states should have a small standard deviation
with respect to a local observable. This is confirmed by Figure \ref{fig:STD} displaying
the standard deviation of the target infidelity when the size of the 
sample increases. Each sample is composed of independent random realizations with different $L$.
We expect convergence as  $\frac{1}{\sqrt n}$, where $n$ is the sample size. 
The standard deviation is quite small for the random phase wavefunction sampling. As expected when random transformation $L$ is larger ($3-RP$ and $6-RP$) the STD decreases.

\section{conclusions}

Laser cooling of the internal degrees of freedom of a molecule is a difficult task due to the large occupied Hilbert space.
A shaped pulse generates a unitary transformation accompanied by spontaneous emission. We optimize the unitary transition 
${\cal U}$ such as after many cooling cycles
the target state is colder than the initial one. 
We have utilized quantum typicality for computing the cooling map. The transformation is calculated with an increasing number of random phase wavefunctions. We show  convergence (of infidelity) to the target state. Doing so while reducing the computational effort.

We have shown that this method can be utilized to model cooling of internal degrees of freedom of molecules. It is anticipated that a series of such transformation can cool the rotation to its ground state.
 
\section{Acknowledgment}
We thank Gerard Meijer for his helpful discussions. Work supported by Israel Science Foundation.

	\bibliography{references}
	\bibliographystyle{ieeetr}
	
\end{document}